\documentclass[pre,superscriptaddress,floatfix,longbibliography,twocolumn]{revtex4-1}

\usepackage{graphics}
\usepackage{graphicx}
\usepackage{amsmath}
\usepackage{amssymb,xcolor}
\usepackage{textcomp}
\usepackage{siunitx}
\usepackage{soul}

\newcommand{\textin}[1]{\mbox{\scriptsize{#1}}}

\definecolor{grisclair}{rgb}{0.6,0.6,0.6}

\newcommand{\beq}{\begin{equation}}
\newcommand{\ee}{\end{equation}}

\begin{document}

\title{Effect of surfactant kinetics on the wetting following the drop impact onto rough surfaces}

\author{S. Rodríguez-Aparicio}
\address{Depto.\ de Ingenier\'{\i}a Mec\'anica, Energ\'etica y de los Materiales and\\ 
Instituto de Computaci\'on Cient\'{\i}fica Avanzada (ICCAEx),\\
Universidad de Extremadura, E-06006 Badajoz, Spain}

\author{M. Herreruela-Rosado}
\address{Depto.\ de Ingenier\'{\i}a Mec\'anica, Energ\'etica y de los Materiales and\\ 
Instituto de Computaci\'on Cient\'{\i}fica Avanzada (ICCAEx),\\
Universidad de Extremadura, E-06006 Badajoz, Spain}

\author{M. G. Cabezas}
\address{Depto.\ de Ingenier\'{\i}a Mec\'anica, Energ\'etica y de los Materiales and\\ 
Instituto de Computaci\'on Cient\'{\i}fica Avanzada (ICCAEx),\\
Universidad de Extremadura, E-06006 Badajoz, Spain}

\author{J. M. Montanero}
\address{Depto.\ de Ingenier\'{\i}a Mec\'anica, Energ\'etica y de los Materiales and\\ 
Instituto de Computaci\'on Cient\'{\i}fica Avanzada (ICCAEx),\\
Universidad de Extremadura, E-06006 Badajoz, Spain}

\author{E. J. Vega}
\address{Depto.\ de Ingenier\'{\i}a Mec\'anica, Energ\'etica y de los Materiales and\\ 
Instituto de Computaci\'on Cient\'{\i}fica Avanzada (ICCAEx),\\
Universidad de Extremadura, E-06006 Badajoz, Spain}

\begin{abstract}
We experimentally analyze the effect of a surfactant on wetting following drop impact on rough surfaces, paying special attention to the role of dynamic surface tension. To this end, we compare the results obtained with Triton X-100, SDS, and Surfynol 465. For concentrations below the critical micelle concentration $c_{\textin{cmc}}$, the evolution of the coverage area is nearly identical for all three surfactants, suggesting that the surfactant concentration is too low to significantly influence droplet spreading. In contrast, pronounced differences emerge due to the distinct dynamic surface tensions of the surfactants at $c/c_{\textin{cmc}}=2$. The evolution of the coverage area during spreading is nearly the same for pure water droplets and those containing Surfynol 465, indicating that surfactant depletion is negligible during the rapid spreading stage. As the Weber number increases, droplet spreading becomes progressively less sensitive to surface tension, thereby reducing the influence of surfactant adsorption kinetics. Nevertheless, Surfynol 465 produces larger coverage areas than Triton X-100 and SDS. The final coverage area is governed by the quasi-static recession of the triple contact line, which is controlled by the receding contact angle. Surfynol 465 consistently yields substantially larger final coverage areas across the range of surface roughness considered in this study.
\end{abstract}
\maketitle

\section{Introduction}
\label{sec1}

The deposition and spread of pesticide droplets on plant surfaces directly affect pesticide utilization rates and efficacy. Pesticides are typically deposited through the impacts of droplets on leaves or other parts of the plant. When a pesticide drop impacts a surface, inertia causes it to spread out into a disc-shaped lamella. Subsequently, surface tension causes the drop to retract. Studies on pesticide deposition must aim to develop methods to maximize wetted leaf surface. Understanding the spreading and retraction of pesticide droplets on leaf surfaces is key to improving pesticide efficiency and reducing contamination. 

The leaf surface is generally rough, which significantly affects droplet impact dynamics. During the spreading stage, roughness can trigger the symmetry break-up, giving rise to finger-like disturbances around the rim. Experience reveals that there exists a critical impact velocity above which the deposited droplet fragments into tiny droplets violently ejected outwards, creating what is known as a splash \citep{Y06,JT16,RG14}. If insufficient kinetic energy has been dissipated, drops can also bounce off the waxy leaf of a plant. When the Ohnesorge number $\text{Oh}_m=\mu/(\rho \sigma_{\textin{eq}} R_{\text{rms}})^{1/2}$ ($\mu$, $\rho$, and $\sigma_{\textin{eq}}$ are the liquid viscosity, density and equilibrium surface tension, and $R_{\text{rms}}$ is the root-mean square roughness of the surface) exceeds the critical value $\text{Oh}_m\approx 0.15$, the splashing velocity threshold decreases as $R_{\text{rms}}$ increases \citep{GBSB21}. This threshold also corresponds to the transition from corona to prompt splashing triggered by surface roughness \citep{GBSB21,QCC19}.

Surface roughness also affects the maximum spreading by modifying surface wettability and contributing to viscous dissipation. The maximum spreading factor $\beta_m=D_{d\text{max}}/D_d$, defined as the ratio of the maximum spreading diameter $D_{d\text{max}}$ to the initial droplet diameter $D_d$, may be significantly reduced by the surface roughness. For instance, an increase in the average roughness $R_a$ from 0.4 $\mu$m to 6.3 $\mu$m decreased by approximately 8\% the maximum spreading factor of water droplets impacting on copper surfaces at $\text{We}\approx 70$ and $\text{Re}\approx 3500$ \citep{CDLL24}, where $\text{We}=\rho v_d^2 D_d/\sigma_{\textin{eq}}$ and $\text{Re}=\rho v_d D_d/\mu$ are the Weber and Reynolds numbers defined in terms of the droplet impact velocity $v_d$. Water droplets impacting on stainless steel surfaces of similar roughness showed a similar reduction for $\text{We}\approx 400$ and $\text{Re}\approx 6800$, but no significant difference for $\text{We}\approx 20$ and $\text{Re}\approx 1500$ \citep{TQWZZ17}. The reduction in the maximum spreading of water droplets impacting grinding stones, compared with that on a flat glass surface, decreased as We increased \cite{LCJH25}. 

The roughness effect on $\beta_m$ depends on the liquid viscosity. In fact, this effect becomes negligible for $\text{Oh}_D=\mu/(\rho \sigma_{\textin{eq}} D_d)^{1/2}\gtrsim 0.05$ \cite{LCJH25}. The chemical nature of the surface and the effect of roughness on its wettability are also important. \citet{SHSD21} studied the influence of the dynamic contact angle for low-roughness ($R_a<1.3$ $\mu$m) steel surfaces and low Weber numbers ($\text{We}<11$). Their results showed the importance of the dynamic contact angle in droplet dynamics after spreading and in the final surface area covered by the droplet. Surface roughness can slow the triple-contact-line recession to the point that this process can be regarded as quasi-static and, therefore, controlled by the (equilibrium) receding contact angle. In fact, leaf wettability is most often assessed on the basis of that contact angle value, which is closely related to leaf structure \citep{PMSBSSRSB19}.

A practical way of enhancing the pesticide droplet deposition on rough leaf surfaces consists of dissolving tiny amounts of surfactants in the liquid \citep{WSJ91,SJLHDWGZHJ17,LFLZLWGLYDWJ21,LMZZGD21,LZWZZGD22,ZYSLGXW23}. Surfactants can affect the droplet impact in several ways. A decrease in surface tension can reduce the advancing and receding contact angles, thereby improving surface wettability. The superiority of certain surfactants in reducing the equilibrium contact angles of water droplets on hydrophobic surfaces remains under investigation. Marangoni stress arises due to the nonuniform surfactant distribution at the interface, which influences the spreading rate by altering the surface flow, the droplet’s shape, and the balance of the forces at the contact line \citep{ZB97,BH20}. Live-oligomeric surfactant molecules aggregate into worm-like micelle networks, which entangle with the surface micro/nanostructures to pin the contact line \citep{LCDFCLYLDLWJ19}. \citet{JWWFW22} have shown that small spherical micelles strongly interact with the micro/nanostructures of the superhydrophobic surface. This, combined with the micelle's rapid molecular diffusion, allows surfactant-laden water droplets to completely deposit and spread on a superhydrophobic surface.

The spreading of droplets on hydrophobic surfaces cannot be adequately described by the equilibrium surface tension alone. The equilibrium surface tension of a surfactant solution is not achieved instantaneously. When a fresh interface forms, surfactant molecules must be diffused/convected from the bulk to the interface and then adsorb onto the surface \citep{ED00,MS20}. The dynamic surface tension (the surface tension of a fresh interface as a function of time as the interface ages) \citep{QSBBBS20} reflects the speed at which these processes occur. When a surfactant-laden droplet impacts a solid surface, a large fresh interface is rapidly created. This suggests that the speed at which the surfactant adsorbs onto the interface affects the spreading phenomenon \citep{FW23,VSQCC24}. Fast-kinetics surfactants are expected to allow efficient wetting \citep{HSGB21}. \citet{LCW20} studied how the surfactant chain length affects the droplet spreading. This has been attributed to the increase in the diffusion speed towards the newly formed interface as the chain length increases \citep{BZGGHHCYZD22}. 

This paper experimentally analyzes the spreading and retraction of surfactant-loaded droplets on rough surfaces. Special attention will be paid to the maximum and final areas covered by the droplet. Understanding the rebound and splash of pesticide droplets on leaf surfaces is relevant to improving pesticide efficiency and reducing contamination. However, we will set this aspect of the problem aside by considering parameter configurations in which these effects do not occur or are insignificant. 

We will consider three surfactants: (i) Triton X-100, a non-ionic surfactant with a low sorption rate, (ii) sodium dodecyl sulfate (SDS), an ionic surfactant commonly used for multiple purposes, and (iii) Surfynol 465, a non-ionic surfactant that offers two key advantages: its ultra-fast adsorption rate and its ability to lower the water-air surface tension to a value below that achieved with SDS.

We will first consider droplet impacts at a moderate Weber number. We will show that, at sufficiently high concentrations, SDS and Surfynol 465 increase the maximum coverage area beyond that achieved with Triton X-100 due to their faster adsorption rates. This increase is similar for both SDS and Surfynol 465. To determine whether Surfynol 465 can enhance droplet spreading above SDS, we will reduce the droplet impact time by increasing the Weber number. We will conclude that droplet spreading is essentially insensitive to the Weber number in this regime; therefore, adding Surfynol 465 confers no advantage. However, Surfynol 465 dramatically increases the final coverage area in both the moderate and high-Weber-number regimes. We attribute this effect to the smaller receding contact angle in the presence of this surfactant.

The paper is organized as follows. We formulate the problem in dimensionless terms and explain our experimental strategy in Sec.\ \ref{for}. Section \ref{sec2} describes the experimental methodology. The results are presented in Sec. \ref{sec3}. The paper closes with some concluding remarks in Sec.\ \ref{sec4}.

\section{Formulation of the problem}
\label{for}

Consider a droplet of diameter $D_d$ impacting on a horizontal surface at the speed $v_d$. The droplet density and viscosity are $\rho$ and $\mu$, respectively, while the liquid-air equilibrium surface tension is $\sigma_{\textin{eq}}$. As mentioned in the Introduction, two relevant dimensionless parameters can be defined in terms of these parameters: the Weber number $\text{We}=\rho v_d^2 D_d/\sigma_{\textin{eq}}$, which measures the ratio of the dynamic and capillary pressures, and the Reynolds number $\text{Re}=\rho v_d D_d/\mu$, which compares inertia and viscosity. 

It should be noted that the Weber number is defined in terms of the equilibrium surface tension rather than the clean-water value, as commonly done in previous experiments. In this way, comparing the results for different surfactants at the same value of We reveals the effect of surfactant kinetics and not the obvious influence of the droplet surface tension before impact. Given the low viscosities of the analyzed liquids, the Reynolds number takes large values ($\text{Re}\sim 10^4$) and is expected to affect droplet dynamics to a lesser extent. However, its influence cannot be completely neglected. The thickness of the disc-shaped lamella after impact is much smaller than the droplet diameter $D_d$, which considerably reduces the effective Reynolds number \citep{SL25}.

In most of our experiments, the droplet is loaded with a soluble surfactant, whose strength is measured by the Marangoni (elasticity) number $\text{Ma}=\Gamma_{\infty}R_g T/\sigma_{\textin{eq}}$, where $\Gamma_{\infty}$ is the maximum packing density, $R_g$ the gas constant, and $T$ the temperature. As shown in Sec.\ \ref{sec2}, the isotherm $\sigma_{\textin{eq}}(\Gamma)$ ($\Gamma$ is the surface density) is similar for Triton X-100, SDS, and Surfynol 465, the three surfactants used in our experiments. This allows us to explain the differences among the droplet impacts in terms of the surfactant kinetics.

We do not consider any model to describe surfactant transport and sorption kinetics, as these become complicated for ionic surfactants such as SDS. Instead, we formally account for this aspect of the problem by introducing the set of parameters $\{{\cal P}_i\}$, whose values are assumed to depend only on the surfactant molecule. Hereafter, the symbols $\{{\cal P}_i\}_{\textin{Tri}}$, $\{{\cal P}_i\}_{\textin{SDS}}$, and $\{{\cal P}_i\}_{\textin{Sur}}$ denote $\{{\cal P}_i\}$ evaluated for Triton X-100, SDS, and Surfynol 465, respectively. The amount of surfactant dissolved in the droplet is quantified through the ratio $c/c_{\textin{cmc}}$, where $c_{\textin{cmc}}$ is the critical micelle concentration.

The substrate effect on droplet impact is characterized by the solid-air $\sigma_{sa}$ and solid-liquid $\sigma_{sl}$ interfacial energies, as well as the effective 3D surface roughness characterized by the arithmetical mean height $\hat{S}_a$. The substrate effect on rapid droplet spreading is characterized by the dimensionless parameter $S_a=\rho v_d^2 \hat{S}_a/\sigma_{\textin{eq}}$, which compares the dynamic pressure $\rho v_d^2$ and the roughness capillary pressure $\sigma_{\textin{eq}}/\hat{S}_a$. Alternatively, one can use the parameter $S_a^*=S_a/\text{We}=\hat{S}_a/D_d$, which is equivalent to $S_a$ for a fixed Weber number. The quasi-static retraction of the droplet is essentially affected by the (equilibrium) receding angle $\theta_r$, which effectively accounts for the influence of $\sigma_{sa}$, $\sigma_{sl}$, and $\hat{S}_a$ on this impact phase.

We analyzed the time evolution of the droplet coverage area $\hat{C}$, i.e., the area of the substrate surface covered by the droplet during the impact. Neglecting the gravity effect,
\begin{equation}
C=C(\text{We},\text{Re};\text{Ma},\{{\cal P}_i\},c/c_{\textin{cmc}};S_a,\theta_r;t),
\end{equation}
where $C=\hat{C}/(\pi D_d^2/4)$ is the dimensionless coverage area and $t$ is the time in terms of the characteristic impact time $t_d=D_d/v_d$. Special attention is paid to the maximum $C_{\textin{max}}$ and final $C_{\infty}$ ($C(t)$ for $t\to \infty)$) coverage areas. It is worth noting that $C=\beta^2$, where $\beta$ is the spreading factor commonly used in previous works.

\begin{figure*}
\begin{center}
\resizebox{0.8\textwidth}{!}{\includegraphics{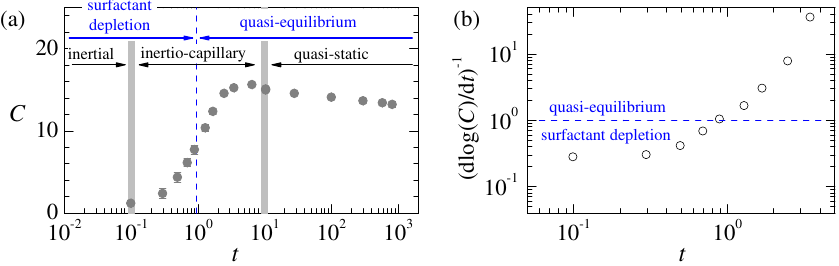}}
\end{center}
\caption{Coverage area $C(t)$ (a) and its growth characteristic time $(d\log C/dt)^{-1}$ (b) for a water droplet loaded with Surfynol 465 impacting on a rough surface ($\text{We}=416$, $\text{Re}=6451$, $c/c_{\textin{cmc}}=1$, and $S_a=1.1$, $\{{\cal P}_i\}_{\textin{Sur}}$ and $\theta_r=23.4^{\circ}\pm 0.3^{\circ}$).}
\label{Mnumer1cc}
\end{figure*}

Figure \ref{Mnumer1cc}a shows the coverage area $C(t)$ for a representative case. One can distinguish three regimes of the droplet evolution. For $t\lesssim 0.1$, the droplet spreading is dominated by inertia and hardly influenced by surface tension. This regime extends over larger values of $t$ as the Weber number increases. For $0.1\lesssim t\lesssim 10$, the liquid dynamics are controlled by both inertia and surface tension (viscosity may also be relevant depending on the surface roughness).  Finally, for $t\gtrsim 10$, the droplet recoils following a quasi-static process. 

We can split the evolution of the surfactant monolayer into two phases. For $t\lesssim 1$, the surfactant does not manage to fully replenish the fresh interface created during droplet spreading. In this interval, the characteristic time of surfactant adsorption is smaller than that of fresh interface creation. The latter can be calculated as the inverse of the interface production rate, $[C^{-1}dC/dt]^{-1}=(d\log C/dt)^{-1}$ (Fig.\ \ref{Mnumer1cc}b). For $t\gtrsim 1$, the surfactant monolayer is essentially at equilibrium. The instant separating these two phases depends on both the surfactant properties $\{{\cal P}_i\}$ and concentration $c/c_{\textin{cmc}}$. 

The surface coverage $C(t)$ depends on the interplay between the impact and surfactant regimes mentioned above. This interplay can be elucidated by comparing the results of a surfactant-loaded droplet with those of a clean-water droplet at the same Weber number (defined in terms of the equilibrium surface tension $\sigma_{\textin{eq}}$). In the surfactant-loaded droplet case, rapid interfacial growth upon impact leads to surfactant depletion, thereby increasing capillary pressure. In addition, the uneven distribution of surfactant on the interface induces Marangoni stress. These two effects are absent in the clean-water experiment. Both increased capillary pressure and Marangoni stress are expected to reduce the droplet's maximum spread with respect to that in clean water. In other words, the decrease in $C(t)$ relative to the clean-droplet case reflects the limited transfer of surfactant to the newly formed interface during the spreading of the surfactant-loaded droplet.

We designed several sets of experiments with different surfactants to determine how the governing parameters influence $C(t)$. First, we conducted experiments with the same equilibrium surface tension. In our experiments, the droplets were ejected by gravitational dripping with the same nozzle. This implies that the droplets had the same diameter $D_d$ and impact velocity $v_d$, independently of the surfactant. Therefore, the dimensionless parameters We, Re, and $S_a$ had the same values for a given substrate. In addition, we verified that $\theta_r$ essentially depends on $\sigma_{\textin{eq}}$ for a given substrate. Under these conditions, the comparison between the experiments with different surfactants allows us to determine the effect of $\{{\cal P}_i\}$ and $c/c_{\textin{cmc}}$ on $C(t)$ (Ma is similar for the three surfactants considered in our analysis, as explained in Sec.\ \ref{sec2}).

We also conducted experiments at the same surfactant concentration $c/c_{\textin{cmc}}$. Interestingly, the small difference in $\sigma_{\textin{eq}}$ between Triton X-100, SDS, and Surfynol 465 is practically compensated by the droplet diameter, so that We and Re took essentially the same values. The difference was minimized by adjusting the droplet velocity. However, the difference in $\sigma_{\textin{eq}}$ yielded significantly different receding angles. This allows us to examine the effect of $\{{\cal P}_i\}$ and $S_a$ on $C(t)$ for small $t$, and $\theta_r$ on $C(t)$ for large $t$.

\section{Materials and methods}
\label{sec2}

\subsection{Rough surfaces}

Translucent rough surfaces were engineered by molding polydimethylsiloxane (PDMS) over grinding papers (SiC). A circular piece of grinding paper, matching the diameter of a plastic Petri dish, was fixed to its base to act as a textured template. A PDMS mixture was prepared at a 10:1 weight ratio (base:curing agent) and thoroughly homogenized. The liquid mixture was then poured onto the template, ensuring complete surface coverage. The system was allowed to cure for 48 hours at room temperature, then subjected to thermal treatment at 70 $^{\circ}$C for 3 hours to ensure complete crosslinking of the polymer. Finally, the PDMS was carefully demolded. 

\begin{figure*}
\begin{center}
\resizebox{1\linewidth}{!}{\includegraphics{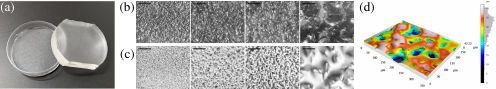}}
\end{center}
\caption{(a) Rough surface fabrication by molding PDMS over grinding papers (SiC). Optical microscope images of (b) grinding papers (SiC) (from left to right: P120, P240, P500, and P1000) and (c) PDMS surfaces. The bar corresponds to 500 $\mu$m. (d) Image of the 3D optical surface profiler for the surface with $\hat{S}_a=7.67$ $\mu$m, which corresponds to P500 paper.}
\label{pdms}
\end{figure*}

Figure\ \ref{pdms}a shows the Petri dish and the resulting rough PDMS substrate, which is translucent due to the remarkable optical transparency of PDMS. This enables visualization of droplet impact from below, which is important given the phenomenon's lack of axisymmetry. Figure\ \ref{pdms}b-c shows the surface of the grinding papers of different grit and the corresponding PDMS substrate observed with a microscope. We used a 3D Surface Metrology Microscope (DCM8) to obtain the topography of the PDMS substrates (Fig.\ \ref{pdms}d) and to evaluate the roughness. Table \ref{T_Rougness} shows the surface arithmetical mean height $\hat{S}_a$ and the root mean square height $\hat{S}_\text{rms}$. In this work, we use $\hat{S}_a$ to characterize the surface roughness. \citet{AZ19} used the same methodology to measure the roughness of some plant leaves, obtaining arithmetical mean height values up to 2.4 $\mu$m.

\begin{table}[]
    \centering
    \begin{tabular}{c|c|c|c}
          & $\hat{S}_a$ ($\mu$m) & $\hat{S}_\text{rms}$ ($\mu$m) \\
        \hline
        P120  & 31.9 & 42.3 \\ 
        P240  & 13.6 & 17.3 \\ 
        P500  & 7.67 & 7.05 \\ 
        P1000 & 4.57 & 5.72 \\ 
    \end{tabular}
    \caption{Grinding papers and PDMS substrate roughness measurements.}
    \label{T_Rougness}
\end{table}

\subsection{Surfactants}

We conducted experiments with Triton X-100 (Sigma-Aldrich), SDS (Sigma-Aldrich), and Surfynol 465 (Evonik).We dissolved the surfactant in ultra-pure Milli-Q water ($\rho=998$ kg/m$^3$ and $\mu=1.0$ mPa$\cdot$s) using a magnetic stirrer for 1 h. All the working solutions were used immediately after production to minimize the effects of surfactant aging and ambient impurities. 

\begin{figure*}
\begin{center}
\resizebox{0.45\linewidth}{!}{\includegraphics{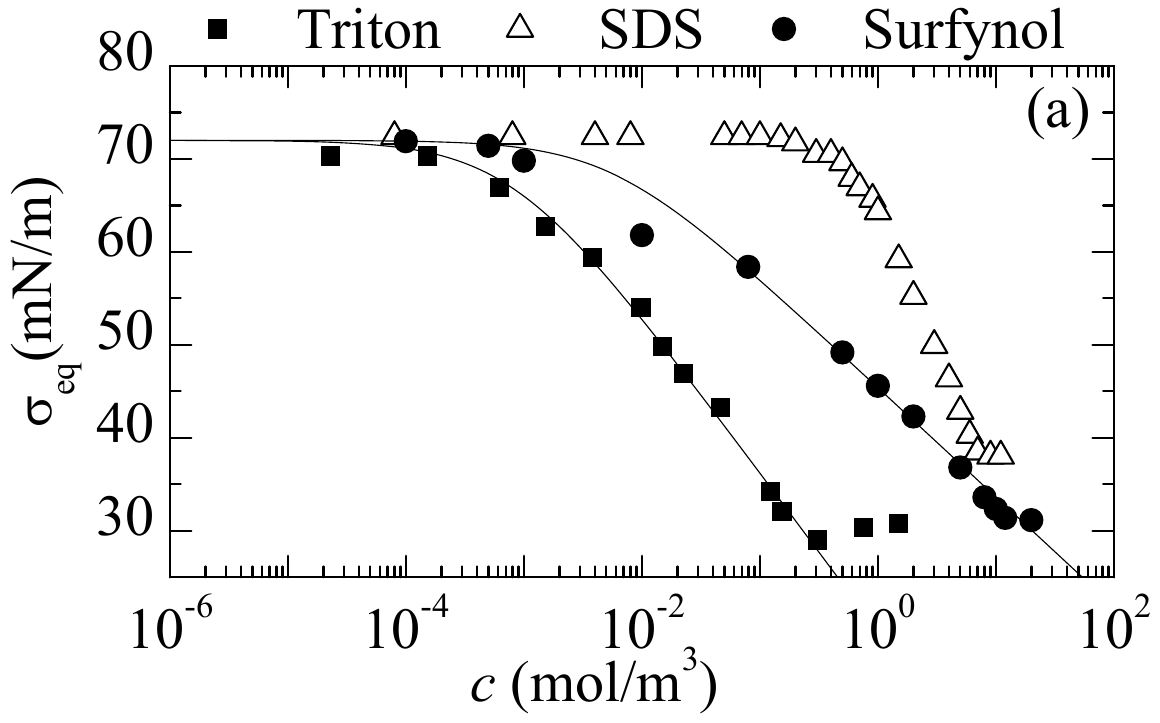}}\resizebox{0.42\linewidth}{!}{\includegraphics{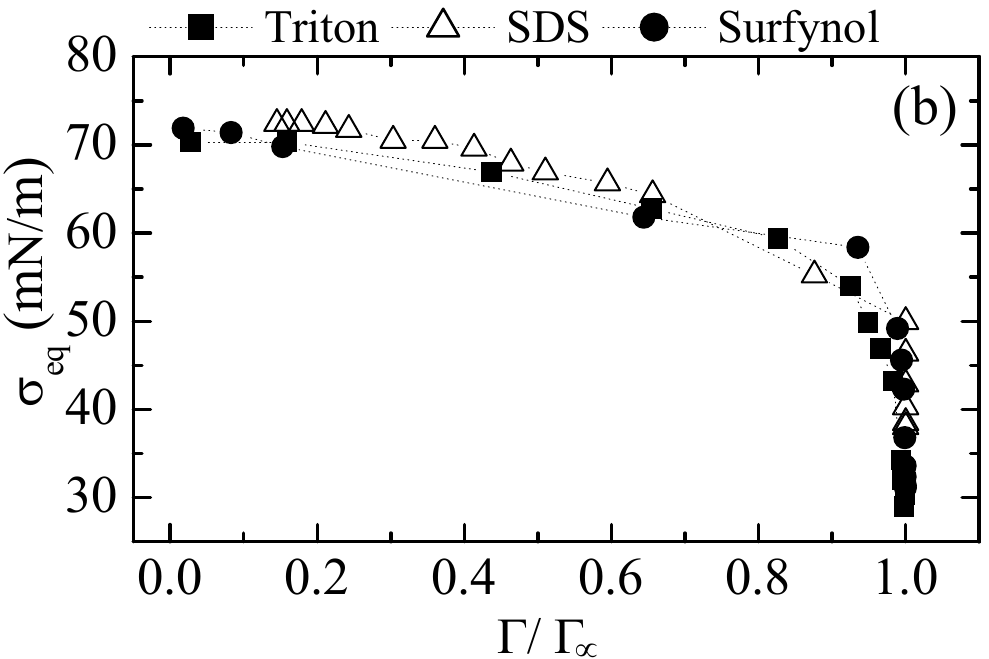}}
\end{center}
\caption{(a) Equilibrium surface tension $\sigma_{\textin{eq}}$ as a function of the surfactant volumetric concentration $c$. The line is the fit of the Langmuir equation of state to the data for Triton X-100 and Surfynol 465. (b) Surface tension $\sigma_{\textin{eq}}$ as a function of the surfactant surface coverage $\Gamma/\Gamma_{\infty}$.}
\label{st}
\end{figure*}

We used the pendant drop method to obtain the equilibrium surface tension $\sigma_{\textin{eq}}$ as a function of the volumetric surfactant concentration $c$ (Fig.\ \ref{st}a). For Triton X-100 and Surfynol 465, the Langmuir equation of state was fit to the experimental data (Figure \ref{st}b). For SDS, we used the measurements of Ref.\ \cite{TMS70}. The critical micelle concentrations are $c_{\textin{cmc}}=0.24$, 8.1, and 10 mM for Triton X-100, SDS, and Surfynol 465, respectively. The maximum packing densities are $\Gamma_{\infty}=3.0$, 3.9, and 2.1 $\mu$mol/m$^2$ for Triton X-100, SDS \citep{TMS70}, and Surfynol 465 \citep{PZPA16}, respectively.

\begin{figure}
\begin{center}
\resizebox{0.88\linewidth}{!}{\includegraphics{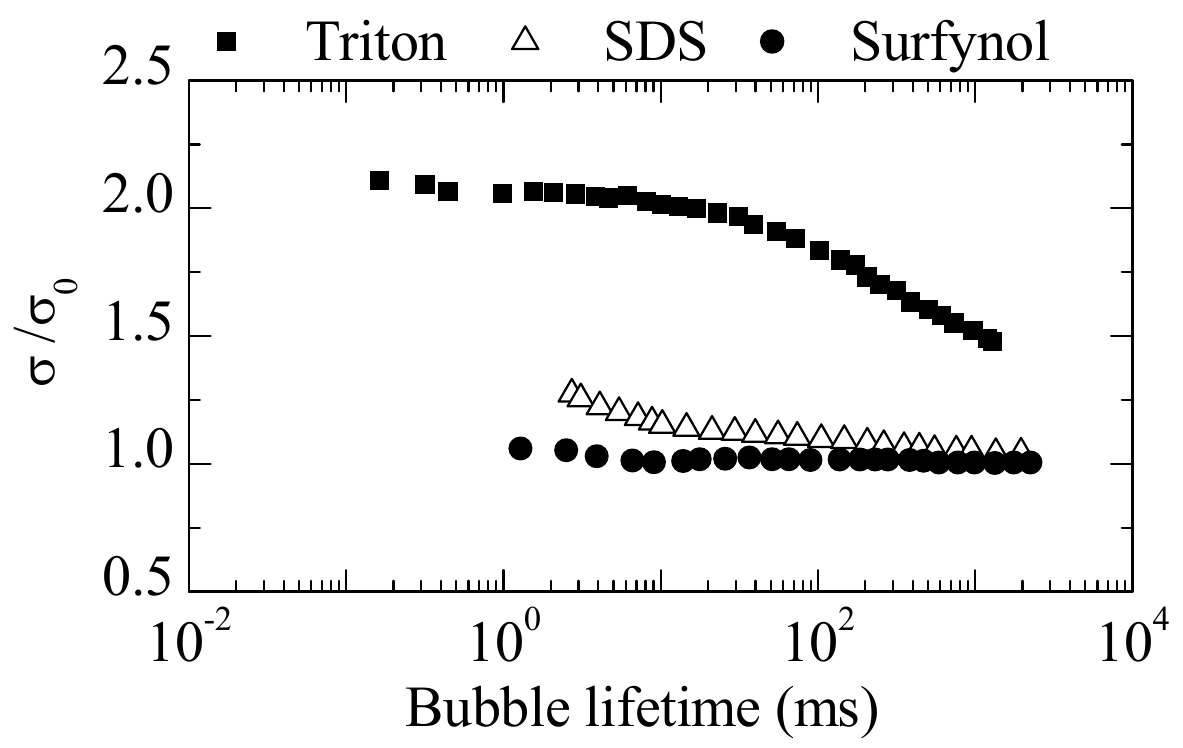}}
\end{center}
\caption{Dynamic surface tension of aqueous solutions of Sufynol 465 at $c/c_{\text{cmc}}=1.3$, SDS at $c/c_{\text{cmc}}=1.3$, and Triton X-100 at $c/c_{\text{cmc}}=1$. The values were normalized with the corresponding equilibrium values. The data was measured by \citet{VSQCC24} using the maximum bubble-pressure tensiometer.}
\label{bmp}
\end{figure}

We chose three surfactants with different sorption dynamics. The results obtained by the maximum-bubble pressure method suggest that a significant transfer of SDS from the bulk to the interface occurs on the timescale of a few milliseconds \citep{VSQCC24}. \citet{VSQCC24} showed that the adsorption rate of Surfynol 465 is even higher than that of SDS at the same relative concentration $c/c_{\textin{cmc}}$ (Fig.\ \ref{bmp}). In this sense, Surfynol 465 is a fast-kinetics surfactant \citep{HLFCM26}. Conversely, Triton X-100 exhibits a much lower adsorption rate in the maximum bubble-pressure tensiometer (Fig.\ \ref{bmp}).

The above statement should be qualified. Figure \ref{bmp} shows $\sigma(\hat{t})$ for essentially the same surfactant concentration $c$ ($c_{\textin{cmc}}=8.1$ and 10 mM for SDS and Surfynol 465, respectively). The rate at which surfactant is transferred to the interface results from two processes: (i) the transfer of monomers from the bulk to the interface sublayer, and (ii) the monomer adsorption at the interface. Above the critical micelle concentration, the first process is driven by convection/diffusion of monomers and micelles toward the interface, and by the disintegration of micelles into monomers near the interface. Assume that, as a result of these transport mechanisms, SDS and Surfynol 465 monomers are transferred to the bubble surface at similar rates in the maximum bubble-pressure tensiometer. Then, one may attribute the differences between $\sigma(\hat{t})$ to the adsorption rate, i.e., the probability per unit time of a monomer to overcome the adsorption energy barrier. However, the number of molecules per unit of area $\Gamma_{\infty}$ that must adsorb to a clean interface to produce the maximum surface tension reduction is almost twice for SDS. Therefore, $\sigma(\hat{t})$ must reach its equilibrium value later for SDS even if the adsorption rate were the same. Comparing dynamic surface tensions of SDS and Surfynol 465 at the same values of $c/\Gamma_{\infty}$ (below the critical micelle concentration) may be a significant measure of the molecules' adsorption rate.

\subsection{Surface wettability}

\begin{figure*}
\begin{center}
\resizebox{0.45\linewidth}{!}{\includegraphics{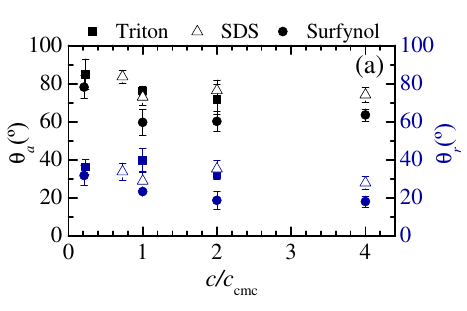}}\resizebox{0.45\linewidth}{!}{\includegraphics{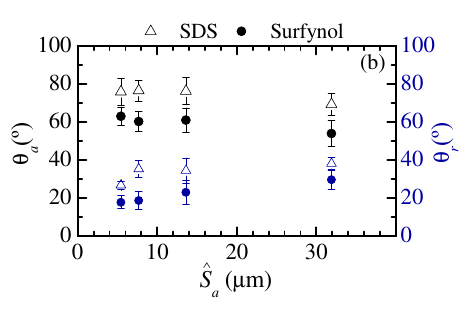}}
\end{center}
\caption{Advancing $\theta_a$ and receding $\theta_r$ contact angle as a function of the surfactant concentration $c/c_{\textin{cmc}}$ for $\hat{S}_a=7.67$ $\mu$m (a), and of the surface roughness $\hat{S}_a$ for $c/c_{\textin{cmc}}=2$ (b).}
\label{angles}
\end{figure*}

Wettability is influenced by surface roughness and by both the type and concentration of surfactant. We measured the (equilibrium) advancing $\theta_a$ and receding $\theta_r$ contact angles with the sessile-drop method \citep{D13b}. Figure \ref{angles} shows the effect of the surfactant concentration and surface roughness on these quantities.
The advancing contact angle is not expected to play a relevant role in droplet spreading \cite{RMT02}. However, as shown in Sec.\ \ref{sec3}, the receding contact angle controls the quasi-static recession of the triple contact line on our rough surfaces. For $c/c_{\textin{cmc}}\geq 2$, $\theta_r$ for Surfynol 465 is significantly smaller than the value for SDS. The maximum difference is found for $c/c_{\textin{cmc}}=2$ and $\hat{S}_a=7.67$ $\mu$m .  

\subsection{Experimental method}

\begin{figure}[tbp]
\begin{center}
\resizebox{0.5\textwidth}{!}{\includegraphics{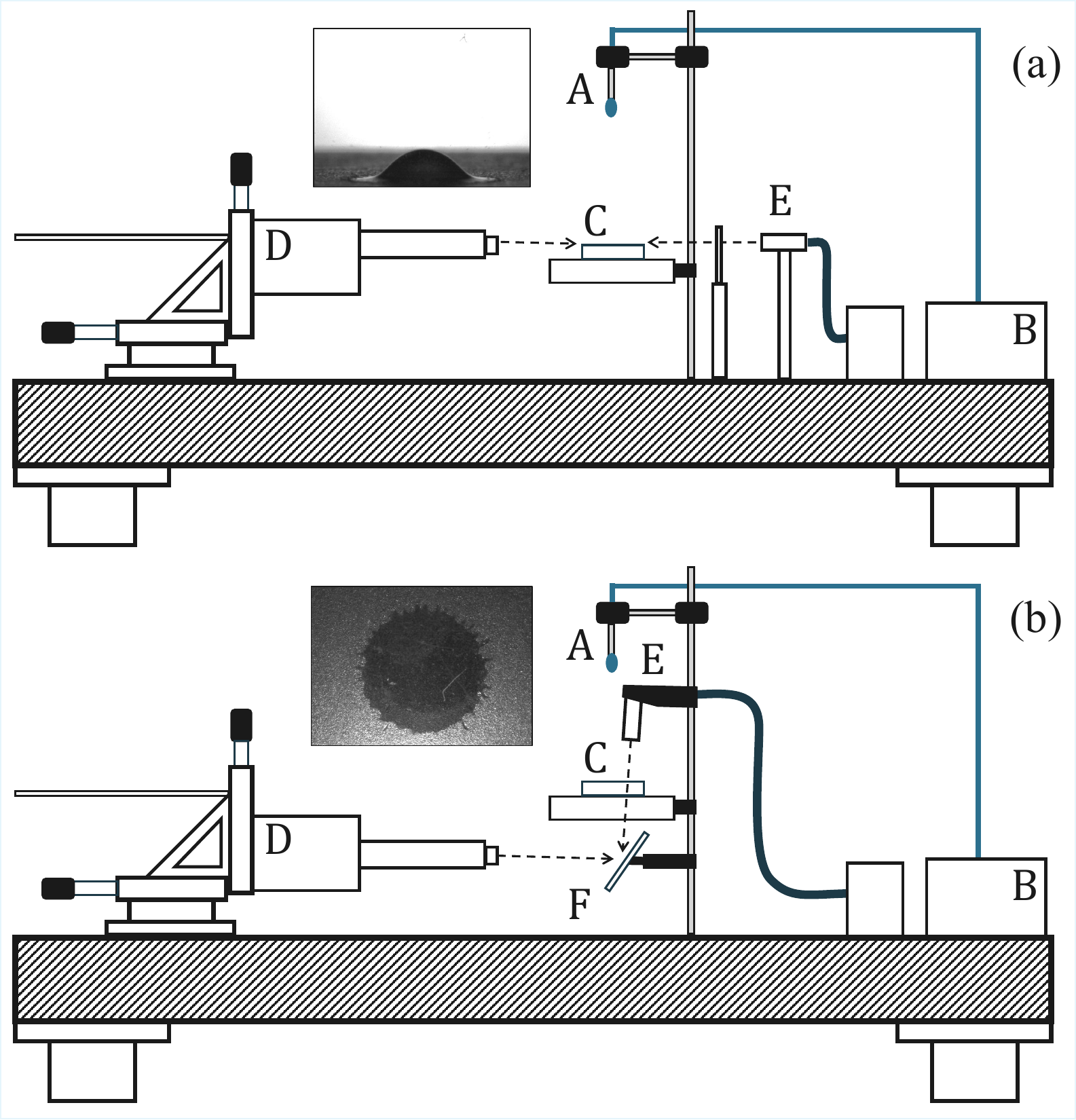}}   
\end{center}
\caption{Experimental setup for capturing the front-view (a) and bottom-view (b) images: feeding capillary (A), syringe pump (B), translucent substrate over a transparent support (C), high-speed camera (D), optical fiber (E), and mirror (F).}
\label{setup}
\end{figure}

The experimental setup is shown in Fig.\ \ref{setup}. The droplet was quasi-statically formed at the tip of the feeding capillary (A), with a radius either 1.27 or 1.8 mm, by injecting liquid via a syringe pump (B). The droplet impacted the rough substrate, which was horizontal on a transparent support (C). The distance between the capillary and the surface was adjusted to select the droplet velocity.  

\begin{figure}
\begin{center}
\resizebox{0.45\textwidth}{!}{\includegraphics{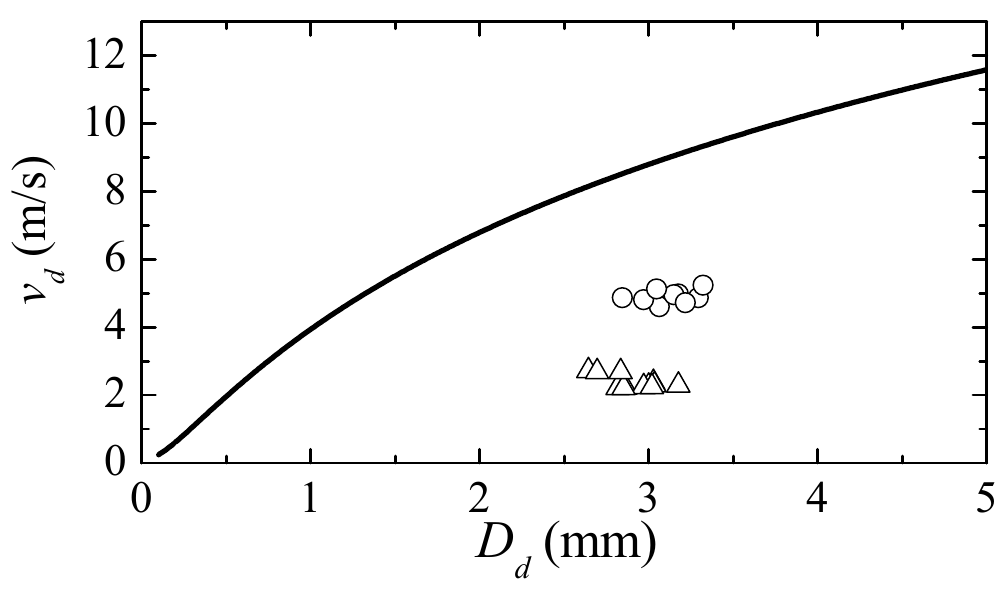}}
\end{center}
\caption{Droplet velocity $v_d$ versus droplet diameter $D_d$ in all the experimental realizations. The triangles and circles correspond to $\text{We}\simeq 400$ and 2000, respectively. The solid line is the terminal velocity for a solid sphere.}
\label{ter}
\end{figure}

\begin{figure*}[tbp]
\begin{center}
\resizebox{0.7\textwidth}{!}{\includegraphics{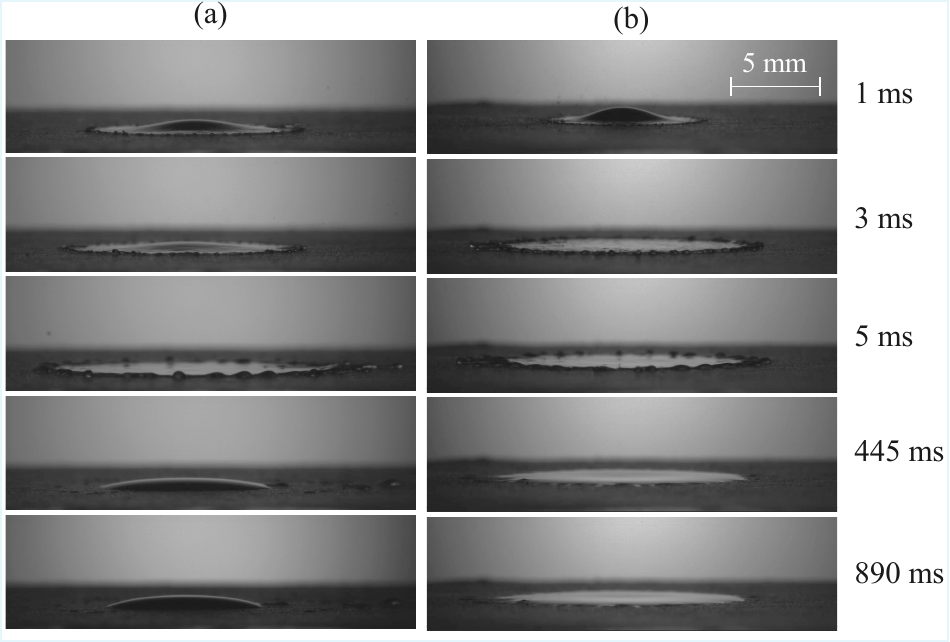}}   
\end{center}
\caption{Sequence of the front-view images of a droplet with SDS (a) and Surfynol 465 (b). The labels indicate the time elapsed after the impact. The experiment corresponds to $\text{We}=2047$, $\text{Re}=14662$, and $c/c_{\textin{cmc}}=2$. The dimensionless roughness is $S_a=5.4$ and 5.6 for SDS and Surfynol 465, respectively.}
\label{front}
\end{figure*}

Front-view images of droplet impact were captured with a high-speed camera (D) and illuminated by an optical fiber (E). In another experimental realization, bottom-view images of the droplet reflected from a mirror (F) were acquired by illuminating the droplet with a vertical optical fiber (E). The front-view images were used to measure its diameter and velocity. As the drop was not spherical, the equivalent diameter was calculated from the volume. Figure \ref{ter} shows the droplet diameter and velocity in our experiments. The velocity was significantly smaller than the terminal velocity. The post-impact front-view images (Fig.\ \ref{front}) allowed us to qualitatively assess the importance of splashing. In most of our experiments, neither the prompt nor the corona splash produced a significant loss of liquid (the ejected volume was less than 5\%).

\begin{figure*}[tbp]
\begin{center}
\resizebox{0.7\textwidth}{!}{\includegraphics{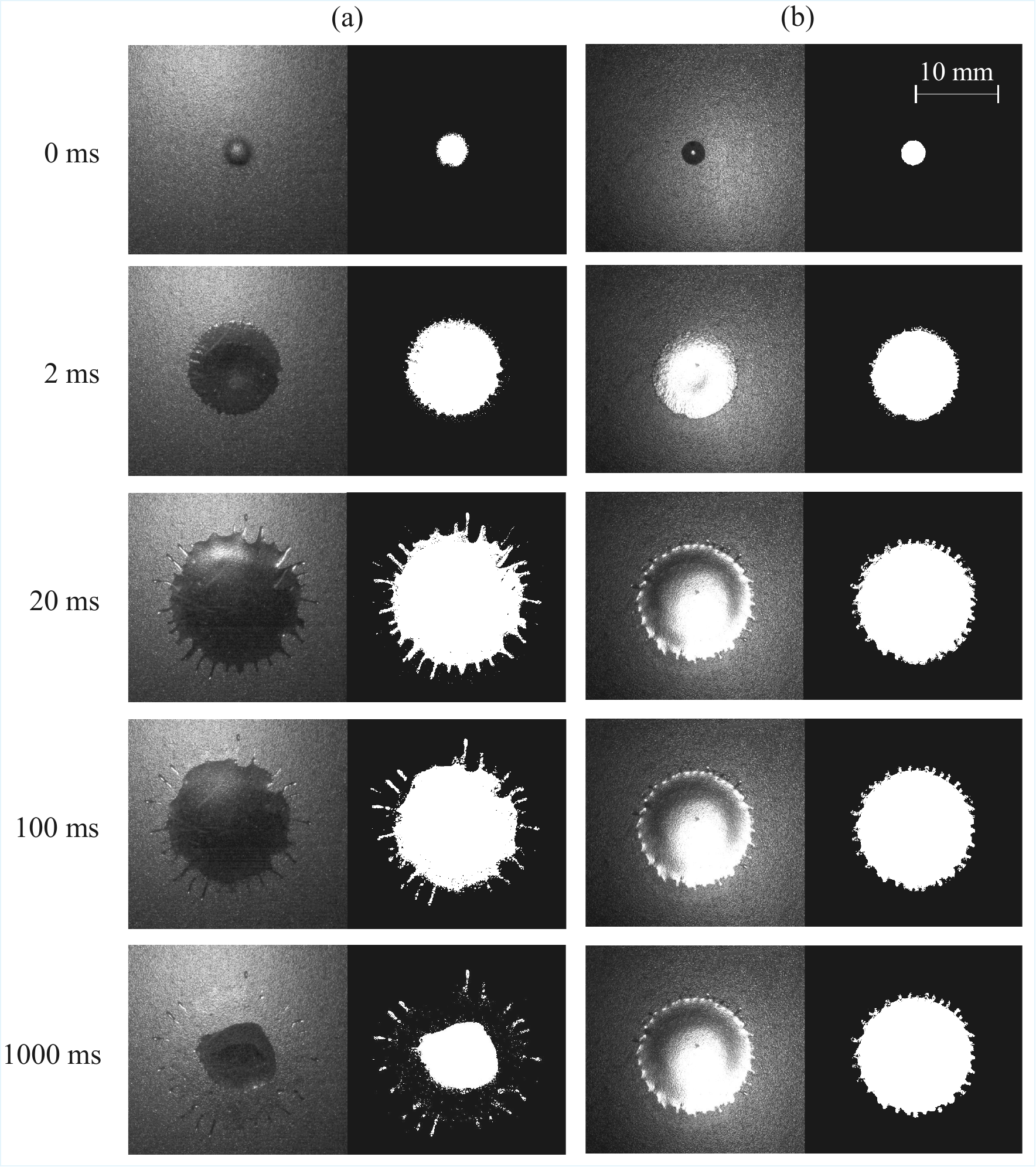}}   
\end{center}
\caption{Sequence of the bottom-view images of a droplet with SDS (a) and Surfynol 465 (b). The figure also shows the processed image. The labels indicate the time elapsed after the impact. The experiment corresponds to $\text{We}=2047$, $\text{Re}=14662$, and $c/c_{\textin{cmc}}=2$. The dimensionless roughness is $S_a=5.4$ and 5.6 for SDS and Surfynol 465, respectively.}
\label{top}
\end{figure*}

Surface roughness hinders the symmetry of the splashing process, especially when working at high Reynolds numbers (Fig.\ \ref{top}). For that reason, we measured the coverage area rather than the spreading diameter, which is typically used for smooth surfaces. The bottom-view images allowed us to calculate the area of the substrate's surface covered by the droplet, excluding splashed (ejected) droplets (Fig.\ \ref{top}). An in-house-developed Matlab routine was developed for this purpose. An image of the background was used to subtract it from the image containing the droplet. Then, a threshold criterion was used to remove background noise and identify the pixels occupied by the liquid. A convolution algorithm was necessary to fill holes inside the droplet. The covered area $A$ was calculated as the whole area covered by the liquid. Most of the area covered by the liquid corresponds to the central circle (Fig.\  \ref{top}). The fingers left behind by the contact line retraction (receding splashing) contributed to that area to a much lesser extent. Therefore, the evolution of the covered area essentially represents the contact line dynamics. The results are discussed in terms of the dimensionless covered area $C=A/(\pi D^2/4)$.

Overall, Figs.\ \ref{front} and \ref{top} show the enhanced wetting produced by Surfynol 465. The surface area covered by the liquid after droplet retraction is considerably larger when SDS is replaced with Surfynol 465, even though the Surfynol-loaded droplet was smaller. We will quantitatively analyze this in Sec.\ \ref{sec3}. Besides, Surfynol 465 partially suppresses receding splashing (droplet breakup as the contact line recedes following maximum spreading). This effect has also been observed when droplets loaded with vesicle surfactants impact superhydrophobic leaves for agriculture \citep{SJLHDWGZHJ17}.

The experiments were repeated 5 times. The results presented in Sec.\ \ref{sec3} correspond to the mean values. The error bars in the figures correspond to the standard deviations. We verified that the results were reproducible on different substrates fabricated with the same grit size. 
  
\section{Results}
\label{sec3}

\begin{figure}
\begin{center}
\resizebox{0.47\textwidth}{!}{\includegraphics{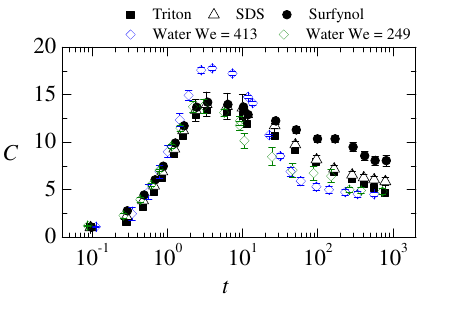}}
\end{center}
\caption{Coverage area $C(t)$ for $\text{We}=413$, $\text{Re}=7121$, $S_a=1.0$, and $\theta_r=33.9^{\circ}\pm 4.8^{\circ}$. The diamonds, squares, triangles, and circles correspond to a clean interface, Triton X-100 ($\{{\cal P}_i\}_{\textin{Tri}}$, $c/c_{\textin{cmc}}=0.25$), SDS ($\{{\cal P}_i\}_{\textin{SDS}}$, $c/c_{\textin{cmc}}=0.73$), and Surfynol 465 ($\{{\cal P}_i\}_{\textin{Sur}}$, $c/c_{\textin{cmc}}=0.21$), respectively. In all the cases, $\sigma_{\textin{eq}}=40$ mN/m. The figure also shows clean-water experiments for $\text{We}=413$, $\text{Re}=9779$, and $S_a=1.0$ (blue diamonds), and $\text{We}=249$, $\text{Re}=7495$, and $S_a=0.6$ (green diamonds). The receding contact angle for the clean-water experiments is $\theta_r=108.4^{\circ}\pm 5.3^{\circ}$.}
\label{Mnumer1}
\end{figure}

First, we examine the droplet impact at a moderately large Weber number $\text{We}\simeq 400$. As explained in Sec.\ \ref{for}, the comparison between the experiments with Triton X-100, SDS, and Surfynol 465 for the same equilibrium surface tension allows us to determine the effect of surfactant adsorption (the dynamic surface tension) on $C(t)$. Figure \ref{Mnumer1} compares the spreading of droplets loaded with Triton X-100 at $c/c_{\textin{cmc}}=0.25$, SDS at $c/c_{\textin{cmc}}=0.73$, and Surfynol 465 at $c/c_{\textin{cmc}}=0.21$. The equilibrium surface tension was $\sigma_{\textin{eq}}=40$ mN/m in the three cases. The droplet coverage rapidly increased for $t\lesssim 3$ ($\hat{t}\lesssim 4$ ms), yielding an increase in the liquid-air interface area of one order of magnitude on that time scale. This phase of droplet spreading may be expected to be sensitive to the surfactant sorption kinetics. However, the evolution of the coverage area is essentially the same across the three surfactants, likely because the concentrations are not high enough for the surfactant to exert a significant effect on the droplet spreading. 

Figure \ref{Mnumer1} also compares the surface coverage of a surfactant-loaded droplet with two clean-water droplet experiments. The first ($\text{We}=249$) corresponds to a clean-water droplet of approximately the same size and velocity as those in the surfactant-loaded droplet cases. This experiment shows spreading without surfactant adsorption, where the surface tension remains equal to that of the clean interface, $\sigma_0$. In the second ($\text{We}=413$), the Weber number is the same as in the surfactant-loaded droplet experiments. The droplet is approximately the same size, and the velocity was adjusted to compensate for the difference in surface tension. If one neglects the effect of the Reynolds number, the spreading in this case corresponds to the limit of ``instantaneous adsorption", i.e., the case in which the surfactant monolayer reaches the equilibrium as soon as the interface is created, and the surface tension takes the equilibrium (minimum) value $\sigma_{\textin{eq}}$ over the whole process. When a surfactant-loaded droplet spreads over the surface, its dynamic surface tension is expected to lie between $\sigma_{\textin{eq}}$ and $\sigma_0$ due to surfactant depletion in the newly created interface. Figure \ref{Mnumer1} shows that $C(t)$ for the surfactant-loaded droplets is essentially the same as that of the clean-water case for $\text{We}=249$. This confirms that the surfactants adsorb little during droplet spreading.

The droplet recession occurs at much longer times, with the surfactant monolayer essentially at equilibrium. The final coverage area $C_{\infty}$ ($C(t)$ for $t\to \infty$) is determined by the quasi-static triple-contact-line recession, controlled by the (equilibrium) receding contact angle $\theta_r$. This angle is similar for SDS and Triton X-100, and is smaller for Surfynol 465 (Fig.\ \ref{angles}). As a consequence, $C_{\infty}$ takes similar values for SDS and Triton X-100, and is larger for Surfynol 465.

From a practical standpoint, it is more relevant to compare results at the same surfactant concentration, at or above the critical micelle concentration, where significant surfactant adsorption is expected during the droplet spreading. As mentioned in Sec.\ \ref{for}, the small difference in $\sigma_{\textin{eq}}$ between Triton X-100, SDS, and Surfynol 465 is compensated by the droplet diameter, so that $\text{We}$ and $\text{Re}$ took essentially the same values in these experiments. However, that difference yielded significantly different receding angles. This allows us to examine the effects of $\{{\cal P}_i\}$ on droplet spreading, and of $\theta_r$ on receding.

\begin{figure}
\begin{center}
\resizebox{0.45\textwidth}{!}{\includegraphics{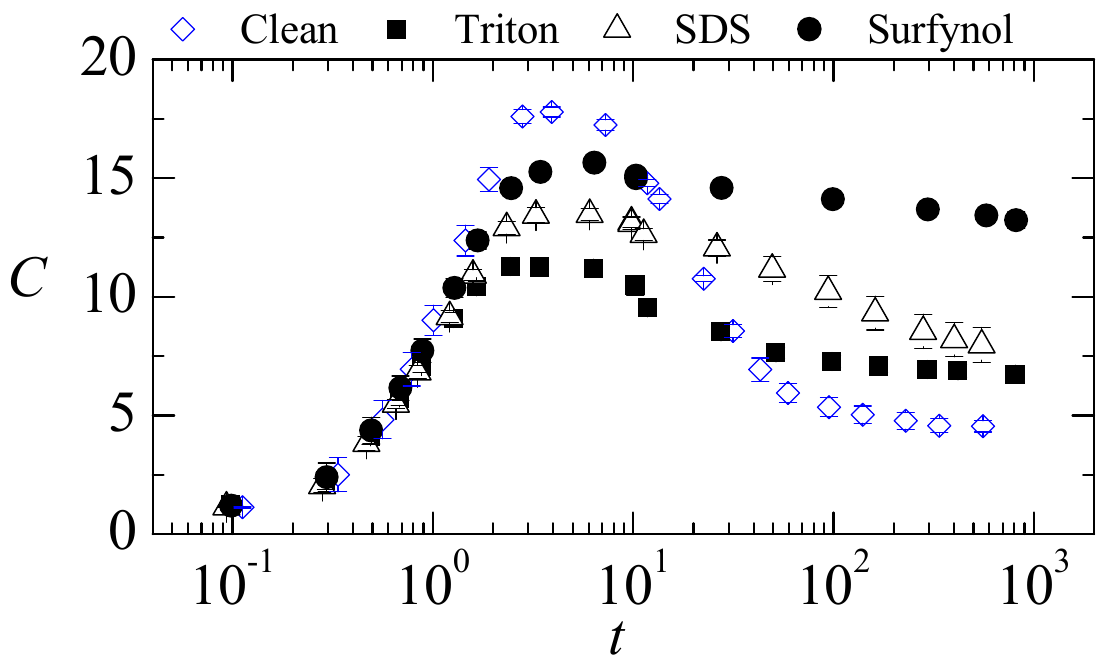}}
\end{center}
\caption{Coverage area $C(t)$ for $\text{We}=416$, $\text{Re}=6451$, $c/c_{\textin{cmc}}=1$, and $S_a=1.1$. The diamonds, squares, triangles, and circles correspond to a clean interface, Triton X-100, ($\{{\cal P}_i\}_{\textin{Tri}}$, $\theta_r=39.8^{\circ}\pm 6.2^{\circ}$), SDS ($\{{\cal P}_i\}_{\textin{SDS}}$, $\theta_r=28.5^{\circ}\pm 4.4^{\circ}$) and Surfynol 465 ($\{{\cal P}_i\}_{\textin{Sur}}$, $\theta_r=23.4^{\circ}\pm 0.3^{\circ}$), respectively. The Reynolds number and the receding contact angle for the water experiments are  $\text{Re}= 9779$ and $\theta_r=108.4^{\circ}\pm 5.3^{\circ}$, respectively.}
\label{numerx}
\end{figure}

When the surfactant concentration equals the critical micelle concentration (Fig.\ \ref{numerx}), significant differences arise during the droplet spreading. The maximum spreading is delayed for SDS and Surfynol 465. It occurs at $t\simeq 6$ ($\hat{t}\simeq 9$ ms). The maximum coverage $C_{\textin{max}}$ qualitatively correlates with the results of the maximum bubble-pressure tensiometer for small bubble lifetime (Fig.\ \ref{bmp}): the larger the dynamic surface tension, the smaller the maximum coverage $C_{\textin{max}}$. However, smaller differences between SDS and Surfynol 465 may be expected given their similar dynamic surface tensions for bubble lifetimes smaller than 10 ms (Fig.\ \ref{bmp}). The maximum coverage for Surfynol 465 approaches that of clean water due to its faster kinetics and higher concentration with respect to that of Fig.\ \ref{Mnumer1}.  However, a significant difference persists, consistent with the higher Reynolds number in the clean-water case \citep{LBBJB14}. In fact, these experiments fall within a parametric region where viscosity plays a significant role on a smooth surface \citep{SL25}. The maximum coverage for Triton X-100 at $c/c_{\textin{cmc}}=1$ is lower than that obtained at $c/c_{\textin{cmc}}=0.25$ (Fig.\ \ref{Mnumer1}), probably due to the Reynolds number effect. Our results do not contradict those obtained by \citet{VSQCC24} for droplet impact on polystyrene. They observed a larger difference in the maximum spreading factor $\beta_{\textin{max}}=C_{\textin{max}}^{1/2}$ of droplets loaded with Triton X-100 and Surfynol 465, as droplet spreading was faster. 

\begin{figure}
\begin{center}
\resizebox{0.45\textwidth}{!}{\includegraphics{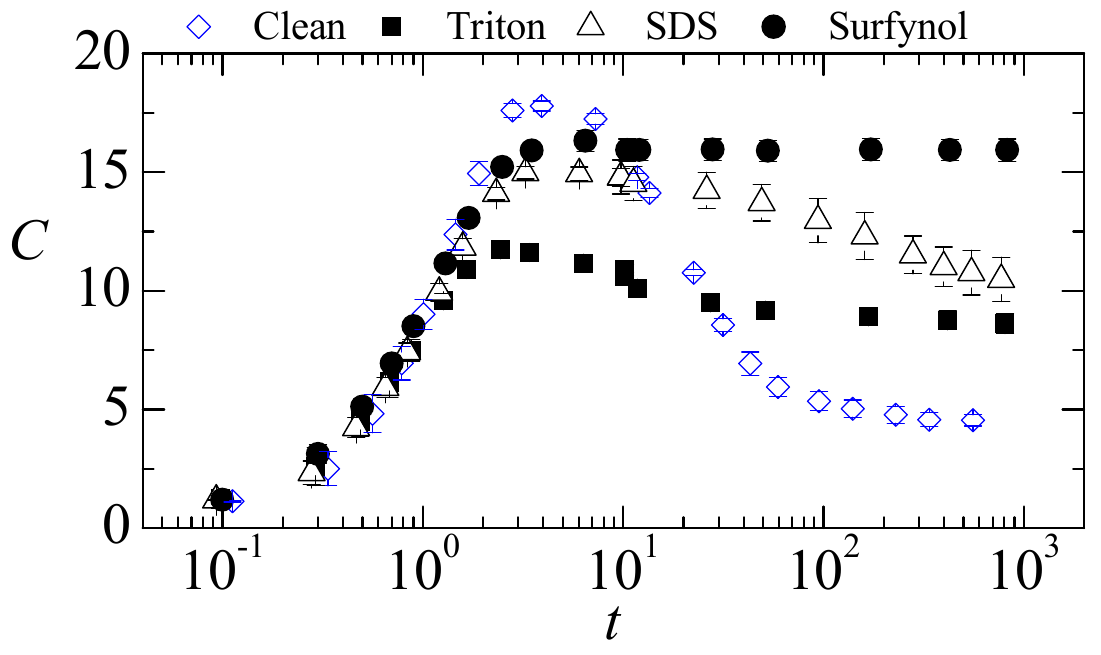}}
\end{center}
\caption{Coverage area $C(t)$ for $\text{We}=416$, $\text{Re}=6477$, $c/c_{\textin{cmc}}=2$, and $S_a=1.1$. The diamonds, squares, triangles, and circles correspond to a clean interace, Triton X-100 ($\{{\cal P}_i\}_{\textin{Tri}}$, $\theta_r=32.3^{\circ}\pm 2.8^{\circ}$), SDS ($\{{\cal P}_i\}_{\textin{SDS}}$, $\theta_r=38.1^{\circ}\pm 3.1^{\circ}$), and Surfynol 465 ($\{{\cal P}_i\}_{\textin{Sur}}$, $\theta_r=18.6^{\circ}\pm 4.7^{\circ}$), respectively. The Reynolds number and the receding contact angle for the water experiments are  $\text{Re}= 9779$ and $\theta_r=108.4^{\circ}\pm 5.3^{\circ}$, respectively.}
\label{numer2}
\end{figure}

Figure \ref{numer2} shows the results for $c/c_{\textin{cmc}}=2$. Both SDS and Surfynol 465 achieve higher maximum coverage than Triton X-100 due to their faster adsorption rates. The mass balance using the impact capillary length leads to $C_{\textin{max}}\sim \text{We}^{1/2}$ for a clean interface and a smooth surface \citep{CBRQ04,JT16}. Interestingly, this scaling approximately verifies if We is calculated with the clean interface surface tension in the case of Triton X-100 and with the equilibrium value in the case of Surfynol 465. This suggests that Triton X-100 adsorbs little during droplet spreading, whereas Surfynol 465 maintains the surface tension relatively constant. In fact, the spreading of the Surfynol-loaded droplet approaches that of the clean-water droplet.

There is a sharp difference between the final coverage areas when Triton X-100/SDS and Surfynol 465 are present. This difference can be attributed to the corresponding receding contact angle values, which govern the quasi-static recession of the triple contact line on the rough surface. Interestingly, the comparison between SDS and Surfynol 465 shows that a 47\% decrease in $\theta_r$ results in a 2-fold increase in the final coverage area, highlighting the critical effect of $\theta_r$.

It has been asserted that the droplet behavior right after impact is primarily determined by surfactants already adsorbed on impact, rather than surfactants adsorbed post-impact \citep{VSQCC24} because the time characterizing this process is too short \citep{RCD22} to allow the adsorption of the surfactants onto the fresh interface generated by the droplet impact \citep{FW23}. This does seem to be the case for the Surfynol-loaded droplet considered in Fig.\ \ref{numer2}. The results suggest that the rapid adsorption of Surfynol 465 maintains the surfactant tension relatively constant throughout the process at $c/c_{\text{cmc}}=2$. This conclusion is consistent with the experiments of \citet{FVMGF26}, who found a negligible reduction in dynamic surface tension during sub-millisecond microjet impact on a liquid pool. 

\begin{figure}
\begin{center}
\resizebox{0.45\textwidth}{!}{\includegraphics{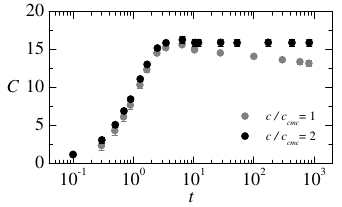}}
\end{center}
\caption{Coverage area $C(t)$ for a droplet loaded with Surfynol 465 at $c/c_{\textin{cmc}}=1$ ($\theta_r=23.4^{\circ}\pm 0.3^{\circ}$) and $c/c_{\textin{cmc}}=2$ ($\theta_r=18.6^{\circ}\pm 4.7^{\circ}$). In both cases, $\text{We}=416$, $\text{Re}\simeq 6460$, and $S_a=1.1$.}
\label{surfynol}
\end{figure}

Figure \ref{surfynol} shows the evolution of the coverage area for two Surfynol 465 concentrations that produce essentially the same equilibrium surface tension. Adsorption is expected to occur for times below $t\lesssim 0.7$ ($\hat{t}\lesssim 1$ ms) (Fig.\ \ref{bmp}). Surfactant adsorbs faster at the highest concentration, reducing the dynamic surface tension and increasing the area. For $t\gtrsim 0.7$, the two droplets are expected to evolve under practically the same dynamic surface tension. The slight difference in maximum coverage reflects distinct dynamic histories of surface tension. Receding dynamics occur quasi-statically and are controlled by the receding contact angle $\theta_r$. They differ markedly between the two cases. For the highest concentration, $\theta_r$ is very small, and almost no receding is observed. This has also been observed for clean water droplets on glass surfaces with similar roughness and wettability \cite{RMT02}.

The adsorption rate of SDS also seems to be sufficiently fast to enhance the droplet spreading. Recent experiments have shown that a millimeter-sized pendant droplet breaks up following essentially the same process in the presence of SDS and Surfynol 465 \citep{RRMH26}. Our results are consistent with these findings. One may wonder whether increasing the Weber number (shortening the droplet impact time) can widen the difference in maximum coverage between SDS and Surfynol 465. To explore this possibility, we conducted experiments for $\text{We}\simeq 2000$. 

\begin{figure}
\begin{center}
\resizebox{0.45\textwidth}{!}{\includegraphics{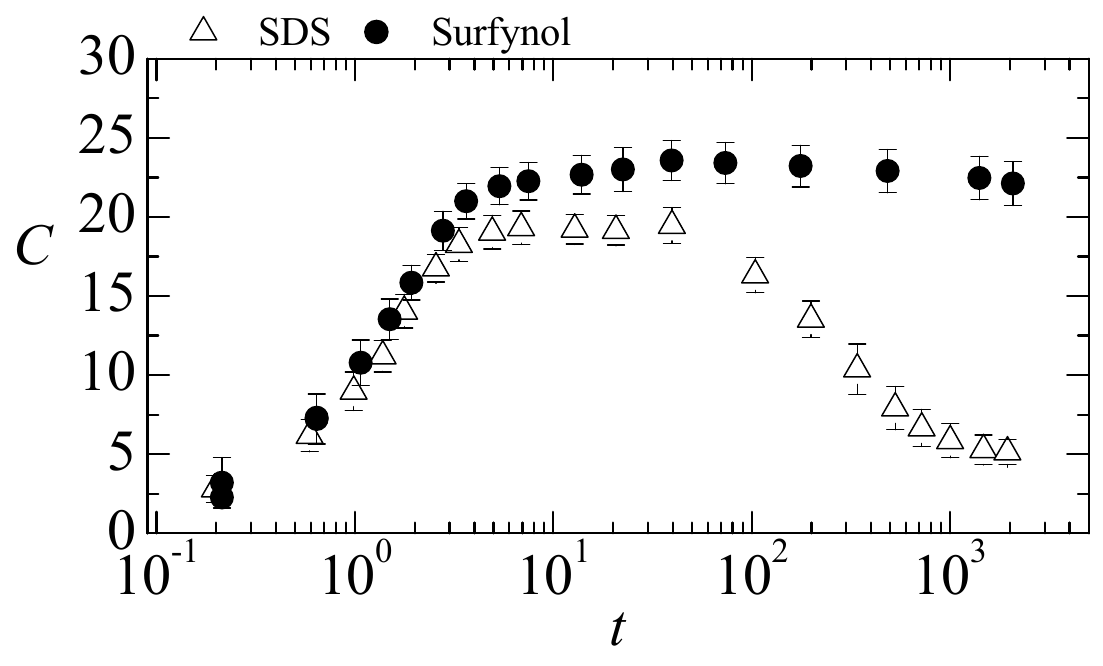}}
\end{center}
\caption{Coverage area $C(t)$ for $\text{We}=2067$, $\text{Re}=14750$, and $c/c_{\textin{cmc}}=2$. The triangles and circles correspond to SDS ($\{{\cal P}_i\}_{\textin{SDS}}$, $S_a=5.4$, $\theta_r=38.1^{\circ}\pm 3.1^{\circ}$) and Surfynol 465 ($\{{\cal P}_i\}_{\textin{Sur}}$, $S_a=5.6$, $\theta_r=18.6^{\circ}\pm 4.7^{\circ}$), respectively.}
\label{numer02}
\end{figure}

Figure \ref{numer02} shows the results for $c/c_{\textin{cmc}}=2$. The coverage area $C(t)$ for Surfynol 465 deviates slightly from that of SDS at $t\sim 10$ ($\hat{t}\sim 6$ ms). Specifically, the maximum value of $C(t)$ for Surfynol 465 is approximately 28\% higher than that in the presence of SDS. This effect is slightly larger than that observed for $\text{We}\simeq 400$, probably due to the reduction of the droplet impact time (the increase in the Weber number).

The droplet spreading becomes practically insensitive to the Weber number as We increases. In fact, \citet{VTSL12} observed a saturation of the maximum spreading of clean droplets on smooth surfaces for $\text{We}\gtrsim 10^3$. In the presence of Surfynol 465 at $c/c_{\textin{cmc}}=2$ (Fig.\ \ref{numer02}), $C_{\textin{max}}\simeq 16$ and 23 for $\text{We}=400$ and 2000, respectively, which implies a significant deviation with respect to the scaling law $C_{\textin{max}}\sim \text{We}^{1/2}$ \citep{CBRQ04,JT16}. This implies that, as the droplet impact time decreases, spreading becomes less sensitive to surface tension and, therefore, surfactant kinetics. This may explain the relatively small differences in droplet spreading between SDS and Surfynol 465, even though their sorption rates differed significantly on the droplet impact timescale $\hat{t}\sim 6$ ms for $\text{We}\simeq 2000$.

As occurs for the $\text{We}=400$, there is a significant difference between the final coverage areas in the presence of SDS and Surfynol 465. This difference can be attributed to the corresponding receding contact angle, which governs the quasi-static recession of the triple contact line. 

\begin{figure}
\begin{center}
\resizebox{0.45\textwidth}{!}{\includegraphics{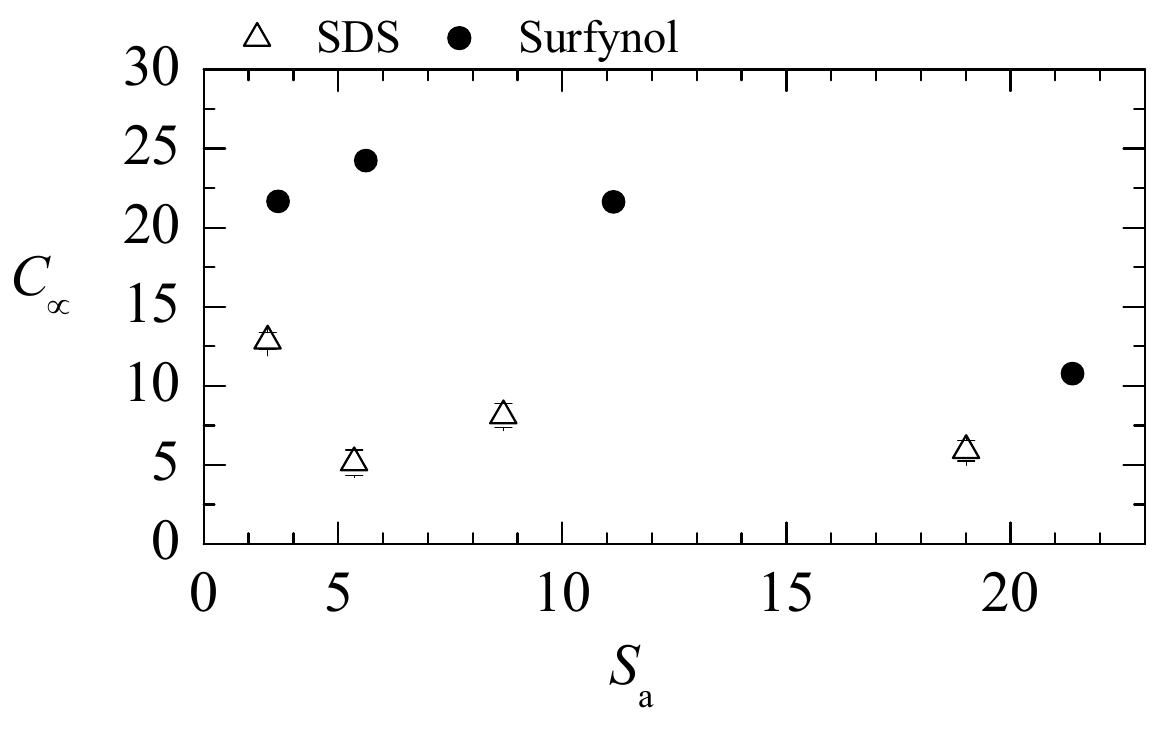}}
\end{center}
\caption{Final coverage area $C_{\infty}$ as a function of the dimensionless surface roughness $S_a$ for $\text{We}=2047$, $\text{Re}=14662$, and $c/c_{\textin{cmc}}=2$. The triangles and circles correspond to SDS ($\{{\cal P}_i\}_{\textin{SDS}}$, $\theta_r=33.6^{\circ}\pm 4.9^{\circ}$) Surfynol 465 ($\{{\cal P}_i\}_{\textin{Sur}}$, $\theta_r=22.2^{\circ}\pm5.4^{\circ}$), respectively.}
\label{numer200}
\end{figure}

The final coverage area $C_{\infty}$ is probably the most relevant quantity in many applications. Figure \ref{numer200} shows the dependence of $C_{\infty}$ on the surface roughness, covering the range of roughness of several typical crop leaves \citep{ZLZHF21,MLDCQZ22}. The receding contact angle for SDS and Surfynol 465 lies in the intervals $26.7^{\circ}\leq \theta_r\leq 38.2^{\circ}$ and $17.6^{\circ}\leq \theta_r\leq 29.6^{\circ}$, respectively (Fig.\ \ref{angles}). This difference explains why $C_{\infty}$ is larger for Surfynol 465, especially for $\hat{S}_a=7.67$ $\mu$m ($S_a=4.5$ and 5.6 for SDS and Surfynol 465, respectively), where the $\theta_r$ deviates the most with respect to the SDS value. For $\hat{S}_a=7.67$ $\mu$m, $C_{\infty}$ increases by a factor of 5 when SDS is replaced with Surfynol 465. In the case for Surfynol 465, $\theta_r$ increases with $\hat{S}_a$ for $\hat{S}_a\geq 7.67$ $\mu$m, which explains the significant decrease in $C_{\infty}$ as the surface roughness increases. The variation of both $\theta_r$ and $C_{\infty}$ is less noticeable for SDS. The non-monotonous behavior of $\theta_r$ on the surface roughness qualitatively explains the non-monotonous dependence of $C_{\infty}$ on $S_a$.

\section{Conclusions}
\label{sec4}

We experimentally analyzed the influence of a surfactant on wetting following drop impact on rough surfaces, paying special attention to the role of dynamic surface tension. For concentrations below the critical micelle concentration, the evolution of the coverage area is essentially the same across the three surfactants, likely because the concentrations are not high enough for the surfactant to exert a significant effect on the droplet spreading. For $c/c_{\textin{cmc}}=2$, significant differences arise across the three surfactants due to their different dynamic surface tensions. The evolution of the coverage area during droplet spreading is essentially the same for water droplets with and without Surfynol 465. Therefore, there is no indication of significant surfactant depletion during the fast droplet spreading. As the Weber number increases, droplet spreading becomes less sensitive to surface tension. This limits the influence of the surfactant kinetics on this phase of the droplet impact. Despite this, Surfynol 465 yields larger coverage areas than those reached with Triton X-100 and SDS. 

The final coverage area is determined by the quasi-static triple-contact-line recession, controlled by the receding contact angle $\theta_r$. The comparison between the results for SDS and Surfynol 465 for $\text{We}=2047$ shows that a decrease in $\theta_r$ of about 47\% leads to a 5-fold increase in the final coverage area. Surfynol 465 leads to considerably larger final coverage areas across the interval of surface roughness considered in our analysis.

Most pesticides begin to be absorbed by a leaf within minutes, but meaningful absorption usually takes about 1–6 hours, depending on the product and conditions \citep{WL07}. In our experiments, the final coverage area $C_{\infty}$ is reached within several seconds. This indicates that pesticide adsorption will be influenced by $C_{\infty}$, rather than the maximum coverage $C_{\textin{max}}$. In this sense, surfactants leading to small receding contact angles, such as Surfynol 465, can enhance pesticide adsorption over the leaf surface. 

\section*{Acknowledgement}

This work was financially supported by the European Union, FEDER and Junta de Extremadura, Autoridad de Gestión (grant. no. IB24089 and GR240077).


%

\end{document}